\documentclass[12pt, onecolumn]{IEEEtran}
\usepackage{amsmath}
\usepackage{amssymb}
\usepackage{mathrsfs}
\usepackage{cite}
\usepackage{epsfig}
\usepackage{epsf}
\usepackage{theorem}
\usepackage{graphics}
\usepackage[active]{srcltx}

\newtheorem{thm}{Theorem}
\newtheorem{lem}{Lemma}

\begin{document}
\renewcommand{\textfraction}{0}

\title{On the Capacity of the Diamond Half-Duplex Relay Channel \footnote{Financial support provided by Nortel and
the corresponding matching funds by the Natural Sciences and
Engineering Research Council of Canada (NSERC), and Ontario Centres
of Excellence (OCE) are gratefully acknowledged.}}
\author{\normalsize
Hossein Bagheri , Abolfazl S. Motahari, and Amir K. Khandani\\
\small Dept. of Electrical Engineering \\[-5pt] \small University of Waterloo \\[-5pt]
\small Waterloo, ON, Canada N2L3G1 \\[-5pt] \small
$\lbrace$hbagheri, abolfazl, khandani$\rbrace$@cst.uwaterloo.ca}
\date{}
\maketitle \thispagestyle{empty}
\begin{abstract}
We consider a diamond-shaped dual-hop communication system
consisting a source, two parallel half-duplex relays and a
destination. In a single antenna configuration, it has been
previously shown that a two-phase node-scheduling algorithm, along
with the decode and forward strategy can achieve the capacity of the
diamond channel for a certain symmetric channel gains
\cite{XueIT07}. In this paper, we obtain a more general condition
for the optimality of the scheme in terms of power resources and
channel gains. In particular, it is proved that if the product of
the capacity of the simultaneously active links are equal in both
transmission phases, the scheme achieves the capacity of the
channel.
\end{abstract}
\normalsize

\section{Introduction}

\emph{Half-duplex} relays i.e., relays that can not transmit and
receive at the same time, have recently attracted enormous attention
due to their simplicity and cost efficiency. Some capacity results
for the case of a single half-duplex relay are presented in
\cite{Host-MadsenIT0105}. To realize multiple-antenna benefits
without increasing the weight and size of the equipments, multiple
relays come into play. A simple model for investigating the
potential benefits of the multiple relays is depicted in Fig.
\ref{fig: model}. The end-to-end capacity of the relay channel has
been studied in \cite{ScheinISIT00, ScheinPhD, XueIT07,
RankovAsil05, ChangAllerton07, SaeedTechrpt08} and is referred to as
\emph{diamond relay channel} in \cite{XueIT07}. Schein in
\cite{ScheinISIT00} and \cite{ScheinPhD} established upper and lower
bounds on the capacity of the full-duplex diamond channel.
Half-duplex relays have been considered in \cite{XueIT07,
RankovAsil05, ChangAllerton07, SaeedTechrpt08}. Xue, and Sandhu in
\cite{XueIT07} proposed several communication schemes including
multihop with spatial reuse, scale-forward, broadcast-multiaccess
with common message, compress-forward, as well as hybrid ones. They
assumed that there is no interference between the transmitting and
receiving relays. However \cite{RankovAsil05,ChangAllerton07,
SaeedTechrpt08} considered such interference and used different
names of \emph{two-way} \cite{RankovAsil05,ChangAllerton07}
 and \emph{successive}\cite{SaeedTechrpt08} relaying, respectively for the \emph{multihop with spatial reuse} protocol. 
In this work, we follow the set-up of \cite{XueIT07} and assume
there
is no such interference. 

In this paper, we are interested in situations where the simple
strategy of successive relaying achieves the capacity of the diamond
channel. Surprisingly, we prove that when the product of the
capacity of the simultaneously active links in both transmission
phases are equal, the scheme achieves the capacity. Note that this
condition includes the result indicated in \cite{XueIT07} as a
special case.

\subsection{Notations}
Throughout this paper,  boldface letters are used to denote vectors.
$A \rightarrow B$ represents the link from node $A$ to node $B$. The
notation also means: \emph{approaches to} when the right hand side
of $\rightarrow$ is a number. A circularly symmetric complex
Gaussian random variable Z with mean 0 and variance $\sigma^{2}$, is
represented by $Z \sim \mathcal{CN}(0,\sigma^{2})$.

\section{Transmission Model over Diamond Relay Channel} \label{sec: System Model}
\begin{figure}[t]
\centerline{\psfig{figure=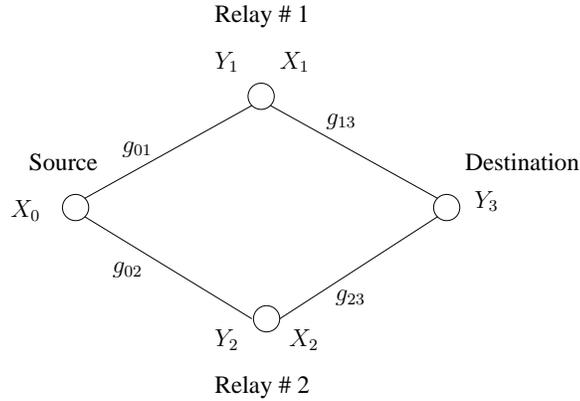,width=3 in}} \caption{The
diamond relay channel} \label{fig: model}
\end{figure}

In this work, a dual-hop wireless system depicted in Fig.\ref{fig:
model}, is considered. The model consists of a source (S), two
parallel half-duplex relays ($Re_{1}$, $Re_{2}$) and a destination
(D). The corresponding index for the nodes are 0, 1, 2, 3, as shown
in the Fig. \ref{fig: model}. We assume that all the nodes are
equipped with a single antenna. Also, due to the long distance or
strong shadowing, no link is assumed between the source and the
destination, as well as between the relays. The background noise at
each receiver is considered to be additive white Gaussian noise
(AWGN). The channel gain between node $i$ and $j$ is assumed to be
constant and is represented by $g_{ij}$ as shown in the Fig.
\ref{fig: model}
. In addition, the channel state information knowledge at different
nodes is as follows
\begin{itemize}
  \item Source knows all the channel gains (\emph{i.e.}, $g_{01}$, $g_{02}$, $g_{13}$ and $g_{23}$).
  \item The relay \emph{i} knows its inward and outward channel gains ($g_{0i}$ and
  $g_{i3}$).
  \item The destination knows its inward channel
gains ($g_{13}$ and $g_{23}$).
\end{itemize}

For this model, the successive relaying scheme is performed in two
stages\footnote{Scheduling is assumed to be done in advance.}:
\begin{enumerate}
  \item In the first $\lambda$ portion of the transmission time\footnote{Total transmission time is T time slots.}, S and $\textrm{Re}_{2}$
  transmit to $\textrm{Re}_{1}$ and D, respectively.
  \item In the remaining $(1-\lambda)$ portion of the transmission time, S and $\textrm{Re}_{1}$ transmit to $\textrm{Re}_{2}$ and D, respectively.
\end{enumerate}

Note that the relays decode and then re-encode what they have
received prior to their transmission. The other possible scheduling
algorithm for the half-duplex diamond channel is called simultaneous
relaying \cite{SaeedTechrpt08} or transmission policy I
\cite{XueIT07} and is as follows:
\begin{enumerate}
  \item In the first $\lambda$ portion of the transmission time, S
  broadcasts its data to both relays.
  \item In the remaining $(1-\lambda)$ portion of the transmission time, relays
  cooperatively transmit to D.
\end{enumerate} It is possible to combine both scheduling methods, however
in this work, successive relaying is considered only. The received
discrete-time complex baseband equivalent signals at $Re_{1}$
$Re_{2}$, and D are respectively given by
\begin{eqnarray}
\label{eq01a} Y_{1}[m] & = & h_{01}  X_{0}[m]+N_{1}[m]\nonumber\\
\label{eq02a} Y_{2}[n] & = & h_{02}  X_{0}[n]+N_{2}[n]\nonumber\\
\label{eq03a} Y_{3}[m] & = & h_{23}  X_{2}[m]+N_{3}[m]\nonumber\\
\label{eq04a} Y_{3}[n] & = & h_{13}  X_{1}[n]+N_{3}[n],
\end{eqnarray}
where $X_{i}$ and $Y_{j}$ are the transmitted and received signals
from and to node $i$ and $j$, respectively (see Fig. \ref{fig:
model}). m$\in \{1,...,\lambda T\}$ and n$\in \{\lambda T+1,...,T\}$
denote the transmission time index corresponding to the two
stages\footnote{It is assumed that $\lambda T$ is an integer.}.
$h_{ij}$ is the complex channel coefficient and is connected to
$g_{ij}$ by $g_{ij}=|h_{ij}|^{2}$. $N_{j}$ is the noise at node j
and $N_{j} \sim \mathcal{CN}(0,\sigma_{j}^{2})$, for $j=1,2,3$. We
assume average power constraint for S, $Re_{1}$ and $Re_{2}$ and
denote the constraints by $P_{S}$, $P_{Re_{1}}$ and $P_{Re_{2}}$,
respectively, \emph{i.e.} for $\lambda T$ tuple and $(1-\lambda) T$
tuple sub-codewords, we have
\begin{eqnarray}
  \frac{1}{\lambda T}\sum_{k=1}^{\lambda T} \mid x_{01}[k]\mid^{2} &\leq& P_{S} \nonumber \\
\frac{1}{(1-\lambda) T}\sum_{k=\lambda T+1}^{T} \mid x_{02}[k]\mid^{2} &\leq& P_{S} \nonumber\\
\frac{1}{\lambda T}\sum_{k=1}^{\lambda T} \mid x_{23}[k]\mid^{2} &\leq& P_{Re_{2}} \nonumber \\
\frac{1}{(1-\lambda) T}\sum_{k=\lambda T+1}^{T} \mid
x_{13}[k]\mid^{2} &\leq& P_{Re_{1}},
\end{eqnarray}
where $x_{ij}[k]$ is the symbol corresponding to the $k^{th}$ index,
in the sub-codeword transmitted by node $i$ to node $j$. Note that
instead of allocating $P_{S}$ to both S $\rightarrow$ $Re_{1}$ and S
$\rightarrow$ $Re_{2}$ links, we could allocate different powers to
them \emph{i.e.}, $\rho P_{S}$ and $\mu P_{S}$ with the condition
$\rho \lambda + \mu (1-\lambda) =1$ for $\rho \geq 0$ and $\mu \geq
0$. However due to the simplicity of implementation and also
exposition, we choose $\rho=\mu=1$. In addition, power budget at the
relay nodes assumed to be fixed over time.

A rate $R$ is said to be achievable for this scheme, if for
$T\rightarrow \infty$, D can decode the message with error
probability $\epsilon \rightarrow 0$.

In the next three sections, we first derive the cut-set upper bound
and the achievable rate using the successive relaying protocol, and
then we state the condition for optimality of the scheme.

\section{Cut-Set Upper Bound} The upper bound for half-duplex
networks is given in \cite{Khojastepour_Cutset} based on writing the
well known cut-set bounds for different transmission states shown in
table I and then adding them up. Tx and Rx denote transmit and
receive modes in the table, respectively.

\begin{table}
   \label{tbl: 1}
\caption{Transmission States in Diamond Relay Channel.} \centering
  \begin{tabular}{|c|c|c|c|}
  \hline
  State & R$_{1}$ & R$_{2}$ & Time\\
  \hline
   $S_{1}$ & Rx & Rx & $t_{1}$\\
  \hline
   $S_{2}$ & Tx & Rx & $t_{2}$\\
  \hline
   $S_{3}$ & Rx & Tx & $t_{3}$\\
  \hline
   $S_{4}$ & Tx & Tx & $t_{4}$\\
  \hline
  \end{tabular}
\end{table}

Applying the procedure to the diamond relay channel, we have
\cite{XueIT07}
\begin{eqnarray*}
R_{C_{1}} &=& t_{1}I(X_{0};Y_{1},Y_{2})+t_{2}I
(X_{0};Y_{2})+t_{3}I(X_{0};Y_{1}){}
\\
&& {}+{}t_{4}.0\\
 R_{C_{2}} &=&
t_{1}I(X_{0};Y_{2})+t_{2}(I (X_{0};Y_{2})+I (X_{1};Y_{3})){}
\\
&& {}+t_{3}.0+t_{4}I (X_{1};Y_{3})\\
R_{C_{3}} &=& t_{1}I(X_{0};Y_{1})+t_{2}.0{}
\\
&& {}+t_{3}(I (X_{0};Y_{1})+I
(X_{2};Y_{3}))+t_{4}I (X_{2};Y_{3})\\
 R_{C_{4}} &=&
t_{1}.0+t_{2}I(X_{1};Y_{3}){}
\\
&& {}+t_{3}I (X_{2};Y_{3})+t_{4}I (X_{1},X_{2};Y_{3}),
\end{eqnarray*}
where $C_{j}$ for $j=1,...,4$ represents the $j^{th}$ cut-set. The
time-sharing coefficients $t_{i}$s satisfy $\sum_{i=1}^{4} t_{i}=1$.
Using Gaussian codebooks and defining
$C_{ij}=\log_{2}(1+g_{ij}\frac{P_{ij}}{\sigma_{j}^{2}})$ with
$\textrm{P}_{ij}$ as the power allocated to that link, we have the
following optimization problem:
\begin{eqnarray}
  \max_{t_{i}\geq 0} R &s.t.& {} \nonumber \\
  R &\leq& t_{1}C_{012}+t_{2}C_{02}+t_{3}C_{01}+t_{4}.0  \nonumber \\
  R &\leq& t_{1}C_{02}+t_{2}(C_{02}+C_{13})+t_{3}.0+t_{4}C_{13}  \nonumber \\
  R &\leq& t_{1}C_{01}+t_{2}.0+t_{3}(C_{01}+C_{23})+t_{4}C_{23}  \nonumber \\
  R &\leq& t_{1}.0+t_{2}C_{13}+t_{3}C_{23}+t_{4}C_{123}  \nonumber \\
  \sum_{i=1}^{4} t_{i}&=& 1,\label{eq: normal_sum}
\end{eqnarray}
where $P_{01}=P_{S}$, $P_{02}=P_{S}$, $P_{13}=P_{R_{1}}$,
$P_{23}=P_{R_{2}}$,
$C_{012}=\log_{2}[1+(\frac{g_{01}}{\sigma_{1}^{2}}+\frac{g_{02}}{\sigma_{2}^{2}})P_{S}]$,
and
$C_{123}=\log_{2}[1+\frac{1}{\sigma_{3}^{2}}(\sqrt{g_{13}P_{Re_{1}}}+\sqrt{g_{23}P_{Re_{2}}})^{2}]$.

\section{Achievable Rate under Successive Relaying}
It has been shown in \cite{XueIT07} (Theorem 4.1) that the maximum
achievable rate under this transmission policy, is given by
\begin{eqnarray}\label{eqnarr: Sum_Rate_DF}
\label{eq: R_SR} R_{SR}&=&\max\{R_{1},R_{2}\}\\
\label{eq: R1_DF} R_{1}&=&\lambda_{1}C_{01}+\min\{\lambda_{1}C_{23},(1-\lambda_{1})C_{02}\}\\
\label{eq: R2_DF}
R_{2}&=&\lambda_{2}C_{23}+\min\{(1-\lambda_{2})C_{13},\lambda_{2}C_{01}\}
\end{eqnarray}
with
\begin{eqnarray}\label{eqnarr: time-sharing-DF}
  \lambda_{1} &=& \frac{C_{13}}{C_{13}+C_{01}} \nonumber \\
  \lambda_{2} &=& \frac{C_{02}}{C_{23}+C_{02}}.
\end{eqnarray}

It has also been proved that the decode and forward scheme is the
best forwarding scheme under the successive relaying protocol
(Theorem 4.2 in \cite{XueIT07}).

In \cite{XueIT07}, it is stated that for the case of
$P_{S}=P_{Re_{1}}=P_{Re_{2}}$, the scheme achieves the cut-set upper
bound if the channel gains are such that $C_{02}=C_{13}$ and
$C_{01}=C_{23}$ (corollary 4.3). Here we generalize the condition of
optimality in theorem \ref{Th1}.

Before proving the theorem, it is noteworthy to find some special
cases in the successive relaying scheme. Lemma \ref{lemma1} states
such cases. In particular, We are interested in the cases where
$R_{1}=R_{2}$ in (\ref{eq: R1_DF}) and (\ref{eq: R2_DF}).

\begin{lem} \label{lemma1} Assuming all links have non-zero capacity, the conditions for
$R_{1}=R_{2}$ are one of the followings:
\begin{eqnarray}
  \label{eq: case1} C_{01}C_{02} &=& C_{13}C_{23} \\
  \label{eq: case2} C_{01} &=& C_{02} \quad \textrm{if} \quad C_{01}C_{02}\leq C_{13}C_{23}\\
  \label{eq: case3} C_{13} &=& C_{23} \quad \textrm{if} \quad C_{01}C_{02}\geq
  C_{13}C_{23}.
\end{eqnarray}
\begin{proof}
See \cite{hossein_1}.
\end{proof}
\end{lem}

The rate in (\ref{eq: R_SR}) associated with each condition of
equality (\ref{eq: case1}-\ref{eq: case3}) is:
\begin{eqnarray}
  R &=& \frac{C_{01}}{C_{13}+C_{01}}(C_{13}+C_{02}) \label{eq: 15}\\
  R &=& C_{02} \\
  R &=& C_{13}.
\end{eqnarray}

Now we want to check whether the special cases stated in (\ref{eq:
case1})-(\ref{eq: case3}) can meet the cut-set bound and hence
achieve the capacity. Our numerical analysis show that the rates
obtained for conditions (\ref{eq: case2}) and (\ref{eq: case3}) can
not meet the bound. Theorem \ref{Th1} states that the first case
indeed meets the cut-set bound and hence gives the capacity of the
diamond channel for the specified channel gains and power resources.

\begin{thm}\label{Th1}
Assuming a diamond relay channel with non-zero capacity links, the
successive relaying scheme achieves the capacity of the diamond
channel if $C_{01}C_{02}=C_{13}C_{23}$.

\begin{proof}

The cut-set bounds can be written as

\begin{eqnarray}
  R_{C_{1}} &=& t_{1}C_{012}+t_{2}C_{02}+t_{3}C_{01}+t_{4}.0  \nonumber \\
  R_{C_{2}} &=& t_{1}C_{02}+t_{2}(C_{02}+C_{13})+t_{3}.0+t_{4}C_{13}  \nonumber \\
  R_{C_{3}} &=& t_{1}C_{01}+t_{2}.0+t_{3}(C_{01}+C_{23})+t_{4}C_{23}  \nonumber \\
  R_{C_{4}} &=& t_{1}.0+t_{2}C_{13}+t_{3}C_{23}+t_{4}C_{123}.  \nonumber \\
\end{eqnarray}
We consider cuts 2 and 3 and try to maximize the
  minimum of these cuts by finding the best time-sharing vector
  $\textbf{t}^{*}\triangleq( t_{1}^{*},...,t_{4}^{*})$.

  \begin{lem}\label{lemma3}
  The time-sharing vector that maximizes the minimum of the cuts 2
  and 3, is obtained by choosing $\textbf{t}^{*}=[0,\frac{C_{01}}{C_{13}+C_{01}},\frac{C_{02}}{C_{02}+C_{23}},0]$, and is in such a way that the cuts give the same rate.


  \begin{lem}\label{lemma4}
  The cut-set rate of cuts 2 and 3 with the time-sharing vector
  $\textbf{t}^{*}$ is $\frac{C_{01}(C_{13}+C_{02})}{C_{13}+C_{01}}$, the same as the rate obtained in (\ref{eq: 15}).
  \end{lem}

  Now assume that we increase $t_{1}^{*}$ and $t_{4}^{*}$ from $0$ to $\epsilon \geq 0$
  and $\eta \geq 0$, respectively. To satisfy the constraint of (\ref{eq:
  normal_sum}), $t_{2}^{*}$ and $t_{3}^{*}$ have to be decreased by $\gamma$ and $\delta$, respectively. Note that one of the $\gamma$
  and $\delta$ can be negative. To hold the condition (\ref{eq:
  normal_sum}), the following relation should exists between the
  adjustments
  \begin{equation}\label{eq: change}
    \gamma + \delta = \epsilon + \eta.
  \end{equation}
  Now let us calculate the rate change of the cuts 2 and 3. By
  considering $C_{01}C_{02} = C_{13}C_{23}$, we have
 \begin{eqnarray}
   \Delta R_{C2} &=& (C_{02}\epsilon  -(C_{02}+ C_{13}) \gamma  + C_{13} \eta)  \label{eq: cut2_rate}\\
   \Delta R_{C3} &=& (C_{13} \epsilon -(C_{02}+ C_{13}) \delta  + \eta)
   \frac{C_{01}}{C_{13}}, \label{eq: cut3_rate}
 \end{eqnarray}
where $\Delta R_{Ci}$ $i=1,...,4$ is the rate difference between the
rate obtained using the modified time-sharing vector and the rate
acquired by the time-sharing vector $\textbf{t}^{*}$.
 Using (\ref{eq: change}) and substituting $\delta$  in (\ref{eq:
 cut3_rate}), we have

 \begin{equation}\label{eq: cut_rate_3}
\Delta R_{C3} = (-C_{02} \epsilon + (C_{02}+C_{13}) \gamma  - C_{13}
\eta)
   \frac{C_{01}}{C_{13}}.
 \end{equation}

Note that the signs of the rate changes in (\ref{eq: cut2_rate}) and
(\ref{eq: cut_rate_3}) are different. Considering the fact that with
the given time-sharing vector in lemma \ref{lemma3}, i.e.
$\textbf{t}^{*}$, we had $R_{C2}=R_{C3}$, but with the adjustments
actually we decrease the minimum of the rates which concludes the
proof.
  It is interesting to see that the optimum time-sharing vector
  $\textbf{t}^{*}$ makes all the cut-set bounds to be equal.
  \end{lem}
Therefore the successive relaying scheme achieves the capacity.
\end{proof}
\end{thm}

\section{Conclusion}\label{sec: conclusion} In this report, the
condition for optimality of the successive relaying scheme in a
diamond-shaped relay channel, has been generalized from
$C_{01}=C_{23}$ and $C_{02}=C_{13}$ to a more general form of
$C_{01}C_{02}=C_{13}C_{23}$.
\section{Acknowledgement}
H. Bagheri would like to thank Mr. Vahid Pourahmadi for helpful
discussions.


\begin{thebibliography}{9}

\bibitem{XueIT07}
F.~Xue and S.~Sandhu,
\newblock ``Cooperation in a half-duplex gaussian diamond relay channel'',
\newblock {\em Info. Theory}, vol. 53, no. 10, pp. 3806--3814, Oct. 2007.

%
%
%
%

\bibitem{Host-MadsenIT0105}
B.~Wang, J.~Zhang, and A.Host-Madsen,
\newblock ``On the capacity of mimo relay channels'',
\newblock {\em IEEE Trans. on Info. Theory}, vol. 51, no. 1, pp. 29--43, Jan.
  2005.

\bibitem{ScheinISIT00}
B.~Schein and R.~Gallager,
\newblock ``The gaussian parallel relay network'',
\newblock in {\em in Proc. IEEE Int. Symp. Inf. Theory}, 2000.

\bibitem{ScheinPhD}
B.~Schein,
\newblock {\em Distributed coordination in network information theory},
\newblock Ph.d., MIT, Cambridge, MA, 2001.

\bibitem{RankovAsil05}
B.~Rankov and A.~Wittneben,
\newblock ``Spectral efficient signaling for half-duplex relay channels'',
\newblock in {\em in Proc. Asilomar Conf. Signals, syst., comput.}, Pacific
  Grove, CA, Nov. 2005.

\bibitem{ChangAllerton07}
W.~Chang, S.~Chung, and Y.~H. Lee,
\newblock ``Capacity bounds for alternating two-path relay channels'',
\newblock in {\em in Proc. of the Allerton Conference on Communications,
  Control and Computing}, Monticello, IL, Oct. 2007.

\bibitem{SaeedTechrpt08}
S.~S.~Changiz Rezaei, S.~Oveis Gharan, and A.~K. Khandani,
\newblock ``Cooperative strategies for half-duplex parallel relay channel:
  Simultaneous relaying versus successive relaying'',
\newblock Tech. {R}ep. UW-ECE 2008-02, University of Waterloo, 2008.

\bibitem{Khojastepour_Cutset}
M.~A. Khojastepour, A.~Sabharwal, and B.~Aazhang,
\newblock ``Bounds on achievable rates for general multi-terminal networks with
  practical constraints'',
\newblock in {\em Inf. Process. Sens. Netw.: Second Int. Work.}, Palo Alto, CA,
  Apr. 2003.

  \bibitem{hossein_1}
H.~Bagheri, A.~S. Motahari, and A.~K. Khandani,
\newblock ``Scheduling for dual-hop communication with half-duplex relays'',
\newblock Tech. {R}ep. UW-ECE 2008-05, University of Waterloo, 2008.
\end{thebibliography}
\end{document}